# Mathematical Modeling for Network Upgrades in Internet Service Provider Infrastructure

Omar M Malallah and Qutaiba I. Ali

*Abstract—* The ongoing growth of the need for superior Internet services creates great pressure on the ISPs as to the accurate estimation of network upgrade need. The following work introduces a mathematical modeling methodology that can assist in assessing network performance and identifying scenarios that require additional investment in network facilities. Bandwidth usage, server load, delay factors and throughput are evaluated for estimating the effect of different traffic conditions on the effectiveness of the network. This paper's simulations prove the model's usefulness in pinpointing such thresholds so that ISPs can prepare their upgrades and prevent performance constrictions on schedule. The results raise awareness of how higher resource efficiency and service provision can be attained through focused resource management and prevention or risk planning. This studies' contribution, thus, is in offering mathematical solutions to the ISPs that they can deploy to manage their operations, address the dynamic customer expectations, and deal with the complexities that continue to characterize the digital environment.

## I. Introduction

THE The complexity of the change process has pressured Internet Service Providers (ISPs) to provide more bandwidth than is actually needed at a specific time. The key to performing network analysis and carrying out upgrades on the network and solving QoS issues and expecting and preventing costly havoc can all be attributed to using up-to-date predictive network analysis. Subscriber usage statistics can be interpreted and analyzed for subsequent modeling to determine when the ISP needs to upgrade their systems with the goal of enhancing clients' satisfaction and reducing operation costs [1]. There is also the ability to assess traffic conditions, behaviors, and technologies at the interpersonal level and at a macro level, through machine learning, simulation techniques, or predictive analytics to facilitate decisions [2].

These predictive strategies are further enriched by factors such as spatial and temporal variation in demand and equipment lifecycle and other factors to estimate future network congestion [3]. Technological advancements like the DTNs help ISPs to model and assess performance in a 'what if' context giving more accuracy, effectiveness and agility than with conventional simulation [4]. With the availability of new computational methodologies, especially graph neural networks, network performance prediction and delay analysis has been implemented in further detail to ensure ISPs are capable of adapting to customer demands and maintaining infrastructure efficiency [5].

## II. Literature review

This section reviews previous works in the field of network performance modeling and upgrade prediction, presented in chronological order to highlight the evolution of methodologies.

Salah et al. (2012) introduced an analytical queueing model for assessing the performance of rule-based network firewalls under normal traffic and DDoS attacks. The model used an embedded Markov chain to evaluate throughput, delay, CPU utilization, and packet loss. The study demonstrated that optimizing frequently activated rules and minimizing the rule base size can prevent significant performance degradation, providing valuable guidelines for firewall resilience under high traffic conditions [1]. Carpio et al. (2021) proposed a traffic forecasting method using Long Short-Term Memory (LSTM) models integrated with a Mixed-Integer Linear Programming (MILP) model and a greedy algorithm. This approach effectively mitigated migrations and replication overhead, reducing operational resource costs by up to 45%, and improved resource allocation in ISP contexts [2]. Hui et al. (2022) criticized conventional simulation methods for being rigid and imprecise, advocating for data-driven approaches like machine learning. These methods, particularly neural networks, improved modeling fidelity, flexibility, and scalability, enabling better performance prediction in Digital Twin Networks (DTNs) while emphasizing the need for unified datasets and modular designs for real-world application [4]. Le et al. (2022) explored how foundational AI models, similar to those used in NLP, could transform networking tasks such as congestion control, anomaly detection, and performance

This paragraph of the first footnote will contain the date on which you submitted your paper for review. It will also contain support information, including sponsor and financial support acknowledgment. For example, "This work was supported in part by the U.S. Department of Commerce under Grant BS123456."

The next few paragraphs should contain the authors' current affiliations, including current address and e-mail. For example, F. A. Author is with the National Institute of Standards and Technology, Boulder, CO 80305 USA (e-mail: author@ boulder.nist.gov).

S. B. Author, Jr., was with Rice University, Houston, TX 77005 USA. He is now with the Department of Physics, Colorado State University, Fort Collins, CO 80523 USA (e-mail: author@lamar.colostate.edu).

T. C. Author is with the Electrical Engineering Department, University of Colorado, Boulder, CO 80309 USA, on leave from the National Research Institute for Metals, Tsukuba, Japan (e-mail: author@nrim.go.jp).

prediction. Despite notable advancements, challenges like data tokenization and the need for comprehensive datasets remain. Initial implementations, like NetBERT, showcased the potential of these AI-driven solutions for networking [6]. Ferriol-Galmes et al. (2022) introduced RouteNet-Erlang, a Graph Neural Network (GNN) model, for network performance prediction. By considering sophisticated traffic patterns and custom scheduling mechanisms, the model significantly outperformed traditional Queuing Theory (QT), achieving a worst-case error of 6% and demonstrating scalability for large, non-Markovian networks [7]. Liu et al. (2023) developed DWNet, a GNN-based model for predicting delay and variation in network performance. By incorporating detailed network geometry and improving message-passing algorithms, DWNet achieved better accuracy and generalization than RouteNet, particularly in unseen network topologies, as measured by Mean Absolute Error (MAE) and Mean Absolute Percentage Error (MAPE) [5]. Downey et al. (2023) addressed the complexity of Next Generation (NextG) network traffic prediction by introducing a "cluster-train-adapt" framework. This approach combined Dynamic Time Warping (DTW) for clustering and LSTMs for prediction, achieving 50% better accuracy compared to traditional methods, and demonstrated adaptability to spatial-temporal variability in traffic [3]. Shen and Li (2023) emphasized the potential of learning-based performance estimators such as RouteNet, MimicNet, and DeepQueueNet for improving network modeling. These tools enhanced delay estimation, throughput prediction, and traffic modeling, addressing limitations of traditional simulation methods in handling scalability and variable topologies [8]. Hui et al. (2023) revisited DTNs, underscoring the efficacy of data-driven approaches, including GNNs and LSTMs, for performance modeling. These methods exhibited superior scalability and precision in network detection, optimization, and dynamic performance analysis, compared to traditional simulators [9]. Skocaj et al. (2023) proposed a predictive latency management framework for 5G networks, using Bayesian Learning (BL) and Graph Machine Learning (GML). By leveraging real-world Key Performance Indicators (KPIs), the framework improved latency forecasting and QoS in complex environments, such as urban vehicular congestion [10]. Malekzadeh (2023) developed 5GMLR, a scalable and modular ML model for predicting KPIs like throughput, delay, and spectral efficiency in 5G networks. The model achieved an average forecast accuracy of 95% and demonstrated flexibility across deployment scenarios, addressing challenges in high-density and varied environmental conditions [11]. Shin et al. (2023) presented a Digital Twin Network (DTN) architecture integrating GNNs and LSTMs for real-time traffic prediction. The approach reduced training time by 25% while maintaining accuracy, providing scalability for applications such as AR, VR, and the metaverse [12]. Tahmasbi et al. (2023) applied formal methods for evaluating network performance by modeling traffic patterns and interactions as logical formulas. This approach identified nuances in fairness, scheduling, and bottlenecks, complementing traditional analytical methods with greater precision in small-scale networks [13]. Qin (2024) introduced HQ-TLP, a GAN-based temporal link prediction model, which combined Graph Convolutional Networks (GCNs) and GRUs to enhance resource allocation and improve prediction accuracy for dynamic networks, such as wireless mesh and data center networks [14]. Nan et al. (2023) proposed a scalable traffic prediction framework for DTNs, leveraging model transfer techniques for adaptability to diverse traffic types. This framework reduced Mean Squared Error (MSE) by 25.8% compared to benchmarks and demonstrated scalability for real-time applications in cellular networks [15]. Panek et al. (2024) developed Configuration-Performance Modeling (CMPM), a neural network-based approach for predicting KPIs like throughput in mobile networks. By incorporating over 100 configuration parameters, the method significantly reduced prediction errors and demonstrated effectiveness in managing large-scale network configurations [16].

### III. METHODOLOGY: A MATHEMATICAL MODELING APPROACH

With growth in technology, specifically on the internet, there is the pressure that comes with meeting the growth in demand for users who expect high performance from ISPs. This is something as important to its ability to anticipate the precise instant and exact fashion in which such upgrades are likely to be required to keep its service offering excellence, cost effective and competitive [17]. This research work presents a mathematical modeling methodology to assess the network performance and determine specific measures that may be required to enhance the network [18], [19].

The model is built to predict call holding time, number of calls, number of dropped calls or packets, and server occupancy against different user traffic patterns. Hence, by extending the parameters as bandwidth, packet size, server capacity, propagational delay, and queue, the model can be viewed significant in analyzing the network characteristics

*A. Key objectives*

• Performance Evaluation: Evaluate the effects of growth in user loads on such parameters as latency, throughput, and resource outlay.

• Upgrade Prediction: Existence of border-line scenarios that mark a capacity shortfall, which would require infrastructure enhancement.

• Scalability Analysis: Devise test scenarios that allow for experimentation, training of forms of network architecture that are larger and more advanced in Bandwidth, load distribution, and queue size for the evaluation of affective upgrades.

The modeling framework integrates a reference system design solution with pre-design bandwidth, server capacity, and propagation profiles. It also advances to an improved plane with more functions, in which specific evaluations of the results derived from both approaches can be made[20]. This systematic approach corrects the problem of ISPs making without concrete data to inform their investment decisions regarding their infrastructure while, at the same time, avoiding unnecessary operational interferences coupled with their associated expenses.

This is so since the model offers ISP's functional recommendations that can be used to proactively strategize on network performance improvements, customer satisfaction, and most importantly resource utilization.

*B. System Overview*

The network architecture modeled in this study is illustrated in Figure 1. It represents a typical hierarchical structure of an Internet Service Provider (ISP) network, designed to serve multiple end-user devices effectively. Key components include:

1. Internet Gateway: The network connects to the broader internet through a gateway, facilitating access to external resources and online services.

2. Web Server: A centralized web server handles incoming client requests, such as web page fetches or data transactions. This server is the focal point of computational and data storage activity.

3. Router: The router acts as the central node, directing data traffic between the web server and the switches that connect to end-user devices. It ensures efficient routing of packets within the network.

4. Switches: Two switches are connected to the router, each serving as an intermediary node to distribute traffic to multiple end-user devices. They play a crucial role in load balancing and preventing packet collisions in the local network.

5. End-User Devices: Each switch connects to a set of end-user devices (e.g., PCs). These devices represent clients generating traffic for the network. Their activities drive the simulation and analysis in this study.

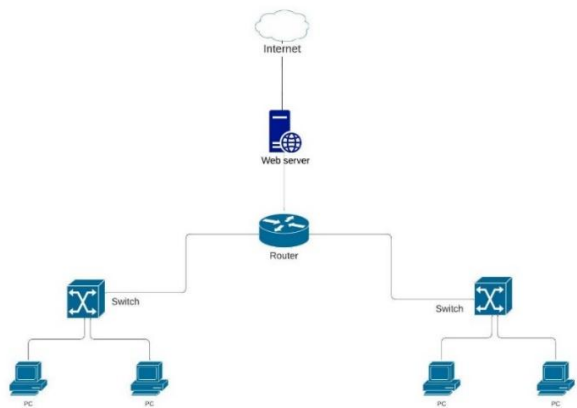

Fig.1: The modelled network architecture.

The model evaluates the performance of this network under varying user loads by incorporating parameters such as bandwidth, server capacity, and queue management. It measures the scalability and efficiency of the system, simulating scenarios where user demands approach or exceed the network's designed capacity. The results guide the identification of upgrade thresholds to maintain service quality.

*C. Assumptions*

- Bandwidth: Fixed bandwidth of B = 100 Mbps.
- Packet Size: Uniform packet size of S = 1500 bytes (12000 bits).
- User Request Rate: For Full HD 1080p video at 30 fps with a maximum bitrate of 5 Mbps, the calculated packet transmission rate is approximately 417 request/second. So $R_{user}$ = 417 requests/second.
- Server Capacity: Maximum server capacity of 50 users.
- Propagation Speed: Fixed propagation speed, v = $2\times10^8$ m/s.
- Average cable length of 90 m.
- Queue Limit = 1000 packets.

*D. Mathematical Formulation*

this research provides an evaluation of network performance with the help of parameters given in the assumption section and the equation which will be discussed in this section. Measures such as transmission delay, propagation delay, queuing delay, throughput, and utilization are used to determine how the network could perform under the existing conditions as well as the enhanced conditions. These equations determine performance limitations and specify values that dictate the need for a network upgrade; they are theoretical benchmarks for improvement.

The equations are grouped into sets based on the structure of the network model to ensure that the structure is comprehensible. The Traffic and Load Modeling equations provided indicate data generation rates and/or arrival activities. Queuing and processing equations relate to queuing and server systems respectively. Delay Components determine transmission, propagation, queuing, and total delay to assess performance. Server Load and Throughput equations evaluate productivity and capability to handle further load, whereas Packet Drops equations focus on situations when more traffic is passing through the limited area than permitted. Upgrade Adjustments include aspects such as higher bandwidth and improved server performance. This organized setup enables a systematic evaluation of the network performance and its further optimization.

1- *Traffic and Load Modelling*

- Traffic Intensity: The traffic intensity $\rho$ is a measure of network load:

$$\rho = \frac{S.R_{total}}{B} \quad (1)$$

Where:
$S$: Packet size (bits).
$R_{total}$: Arrival rate (requests per second).
$B$: Bandwidth (bits per second)

- User Request Rate: The total request rate (request/second) generated by N users:

$$R_{total} = R_{user}.N \quad (2)$$

2- *Delay Components*

- Queuing Delay: The queuing delay depends on traffic intensity $\rho$:

$$D_{queue} = \begin{cases} \frac{\rho}{1-\rho}, & \rho < 1 \\ \infty, & \rho \geq 1 \end{cases} \quad (3)$$

- Processing Delay: The time taken by the server to process requests:

$$D_{processing} = \frac{R_{served}}{server_{capacity}} \quad (4)$$

- Transmission Delay: The time required to transmit a packet over the link:

$$D_{transmission} = \frac{S}{B} \quad (5)$$

- Propagation Delay: The time taken for data to travel over a physical link of average length L:

$$D_{propagation} = \frac{L}{v} \quad (6)$$

- Total Delay: The overall delay experienced by a packet:

$$D_{Total} = D_{propagation} + D_{transmission} + D_{queue} + D_{processing} \quad (7)$$

3- *Server Utilization*

- Server Utilization: The percentage of server capacity being utilized:

$$Server\ Utilization = \frac{R_{served}}{server_{capacity}} \times 100 \quad (8)$$

4- *Throughput*

- Throughput: The effective data rate of successfully transmitted packets:

$$Throughput = \frac{R_{served} \times S}{D_{total}} \quad (9)$$

5- *Packet Drops*

- Queue Drop: Packets dropped due to exceeding the queue limit:

$$Q_{dropped} = max\ (0, min\ (R_{total} - \frac{B}{S}, Queue\ limit)\ ) \quad (10)$$

- Server Drop: Requests dropped due to server overloading:

$$R_{served} = min\ (R_{total}, server_{capacity}) \quad (11)$$
$$server_{drops} = R_{total} - R_{served} \quad (12)$$

6- *Threshold Identification for Upgrades*

Define the criteria for requiring a network upgrade. Examples:

- Total Delay: Identify a maximum acceptable delay based on application requirements. For example: Real-time applications (e.g., video streaming) may require delay <100 ms.
- Define a minimum acceptable throughput e.g., 95% of the theoretical maximum or 5 Mbps for Full HD streaming per user.

7- *Upgrade Scenario Adjustments*

- Upgraded Bandwidth: In the upgraded scenario, B is increased (e.g., from 100 Mbps to 1 Gbps).

- Queue Scaling: The upgraded queue is
- scaled by 5 (for example).

*E. Simulation Program*

The network equations in this study were simulated using Python, chosen for its flexibility and efficiency in scientific computing. The program integrates the mathematical formulas for traffic modeling, queuing dynamics, delays, throughput, and utilization.

Key features include:
- Parameterized Inputs: Adjustable network parameters such as bandwidth, server capacity, and user load.
- Baseline and Upgraded Scenarios: Simulates both the original and enhanced network configurations.
- Data Visualization: Generates graphical outputs for trends like delay and throughput.
- Result Storage: Outputs results in structured formats like Excel and CSV for analysis.

This program enables efficient evaluation of network performance, comparison of configurations, and identification of upgrade thresholds.

## IV. Results and Discussion

The analysis of network performance under the original and upgraded scenarios provides valuable insights into the limitations of the original configuration and the significant improvements achieved through network upgrades. The findings are supported by Figures 2 through 10, which highlight key performance metrics before and after the upgrades.

In the original configuration (Figures 2–5), the total delay per user remained negligible for user counts below 30, as shown in Figure 2. However, beyond this threshold, the delay escalated sharply, reaching a peak of 2 seconds per user. This indicates the network's inability to manage higher traffic loads effectively, leading to severe latency issues that could disrupt real-time applications. Figure 3 illustrates the throughput behavior, where it peaked at approximately 80 Mbps for user counts between 15 and 20, but dropped sharply afterward. For user counts exceeding 30, throughput collapsed entirely, demonstrating that the system was overwhelmed and unable to process additional data efficiently.

Queue drops in the original network, depicted in Figure 4, were minimal up to 30 users. Beyond this point, the number of dropped packets increased rapidly and stabilized at the queue limit of 1000 packets, reflecting inadequate queue capacity to handle high traffic. Server utilization (Figure 5) reached 100% at around 15 users, leaving no room for additional traffic. This full saturation of the server caused severe bottlenecks and significantly contributed to the observed declines in network performance.

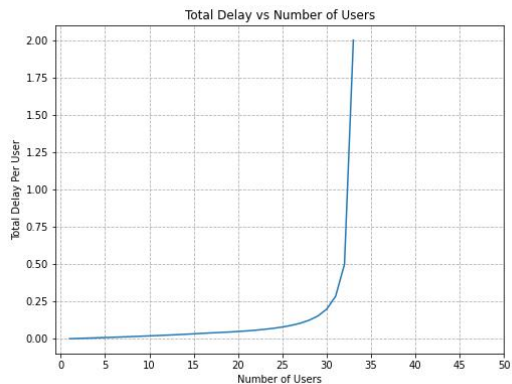

Fig. 2: Total Delay vs Number of Users.

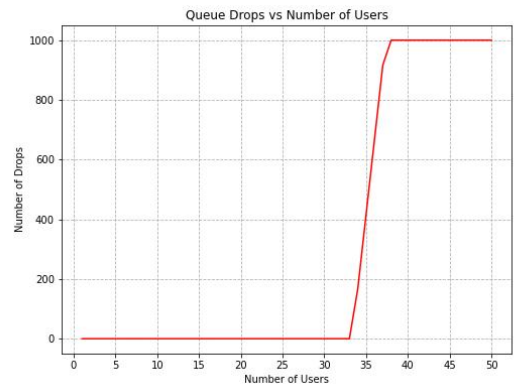

Fig. 4: Queue Drops vs Number of Users.

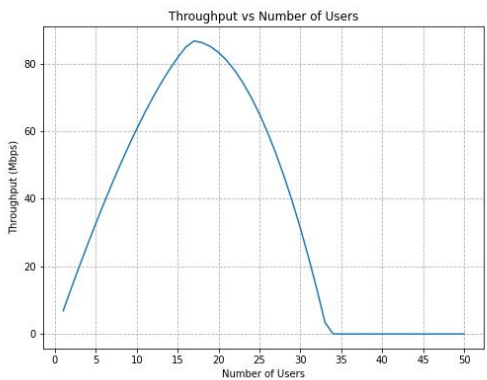

Fig. 3: Throughput vs Number of Users.

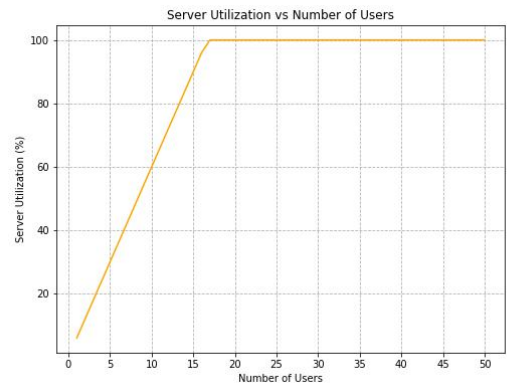

Fig. 5: Server Utilization vs Number of Users.

After implementing the network upgrades (Figures 6–9), substantial improvements were observed. Figure 6 shows that the total delay per user stabilized at approximately 0.02 seconds, even as the user count reached 50. This demonstrates the upgraded network's ability to manage higher traffic loads without latency issues, ensuring consistent performance across all levels of demand. Throughput, as illustrated in Figure 8, improved dramatically, increasing linearly with the number of users and reaching over 400 Mbps at 50 users. Unlike the original network, the upgraded configuration eliminated the throughput bottlenecks, reflecting the enhanced bandwidth and server capacity.

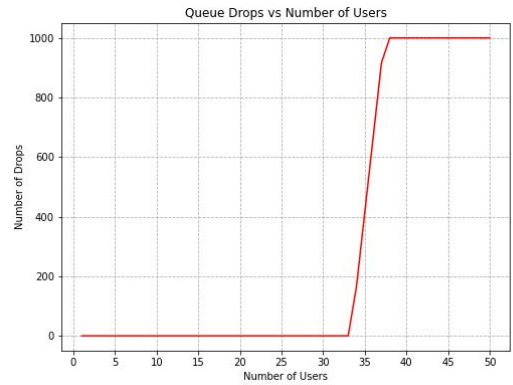

Fig. 8: Queue Drops vs Number of Users.

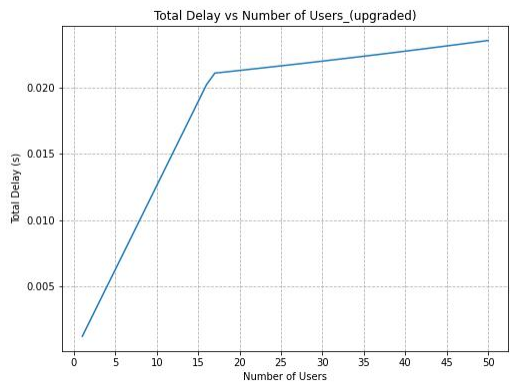

Fig. 6: Total Delay vs Number of Users.

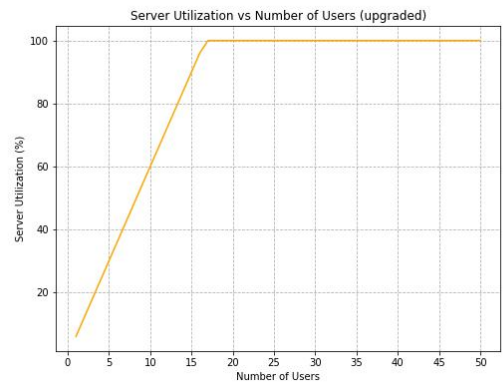

Fig. 9: Server Utilization vs Number of Users.

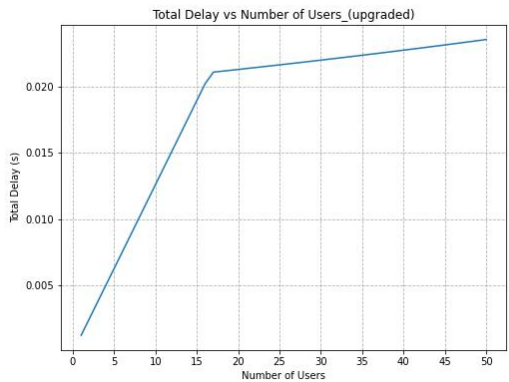

Fig. 7: Throughput vs Number of Users.

Queue drops were entirely eliminated in the upgraded network, as evidenced in Figure 8. Even as the user count reached 50, the system managed all incoming traffic without packet loss, showcasing the success of the upgraded queue management system. Finally, server utilization in the upgraded network, shown in Figure 9, followed a similar trend to the original network, reaching 100% at around 15 users. However, the upgraded system's enhanced bandwidth and reduced delays offset the impact of full utilization, enabling the network to maintain high throughput and low latency under heavy traffic conditions.

## V. CONCLUSION

The constantly growing number of users and evolving nature of network applications and services require accurate and timely prediction of the network's performance and future development needs in ISP networks. This study developed an analytical modeling framework to assess the performance of the network based on the traffic-demand levels and establish the traffic limits that need upgrading of the network system. Combining bandwidth, server capacity, delay components, and throughput, ISPs have a valuable model that translates current limitations into precise goals and analyses for an upgrade.

The overall simulation analysis proves the effectiveness of the proposed approach to identify and prevent performance degradation in the original and enhanced design. The updated network model addresses and demonstrates how ISPs can scale their resources to support much larger numbers of users at the same level of service quality. The benchmarks of servers like utilization, delay and packet loss were observed and explained the need to make investments in bandwidth, server capacity and queue management to facilitate traffic increase in future.

This research highlights the need for Data Analysis techniques and complex mathematical models for the improvement of infrastructural designs of the networks. With the proposed approach ISPs get the practical directives as how to better serve their customers, where to cut costs, which areas need investment for future development. The outcomes are target-based for the tactical network upgrading providing a guideline on how to optimize the network advancement to enhance the service delivery within growing diverse digital environment.